# What is next for National Meteorological Services?

Arribas, A. * [1,2], Robinson, N.H.[1,2], and Evans, P.[1]

1 Met Office Informatics Lab, Exeter, Science Park, EX5 2FS, United Kingdom

2 University of Exeter, Exeter, EX4 4PY, United Kingdom

* Corresponding Author:

Alberto Arribas

Met Office Informatics Lab, Upper Richardson, Science Park, Exeter EX5 2FS, UK

alberto.arribas@informaticslab.co.uk

# What is next for National Meteorological Services?

ABSTRACT

Weather prediction is one of the greatest scientific and technical successes of the last century. Today, National Meteorological Services routinely produce forecasts that have a major positive impact across society and many economic sectors.

This success has been intrinsically linked to improvements in supercomputing which has increased various orders of magnitude following Moore's Law, enabling underpinning increases in model complexity and resolution. However, there are good reasons to believe that we may have reached the end of this particular road.

The combination of technological discontinuities and contextual changes mean that NMS are facing the highest level of uncertainty and change in many decades. This brings new organisational challenges, altering existing power and social structures within NMS. Analysis from other industries that have faced similar transformations in the past show that the period of change we are entering could be as long as 30 years and that there is a substantial risk that the foundations of the weather industry could be altered significantly.

Therefore, NMS need to use their resources not only to make best use of the diminishing improvements available within the current technology trajectory but to simultaneously innovate in new technologies to ensure they can generate value in the future.

This paper analyses the strategic options available to National Meteorological Services.

# What is next for National Meteorological Services?

**Introduction**

Weather prediction is a remarkable scientific achievement. While political, economic or sport forecasts have not got any better over the last few decades, meteorology has significantly improved its ability to predict the future. Despite the jokes, it seems that "The weatherman is not a moron" is still an apt headline[1].

As described in Bauer et al. (2015), during the last seventy years the skill of weather forecasts has increased at a rate of approximately one day for every ten years of R&D (i.e. today's forecast 3-day ahead is as good as a 2-day forecast was ten years ago). This success is a direct consequence of the accumulation of decades of unglamorous scientific research and technological development which has enabled National Meteorological Services (NMS) across the world to routinely produce weather forecasts which are incredibly sophisticated - computationally as complex as simulating the evolution of the Universe - and, in addition, widely useful, being fundamental for decision-making in industries as varied as transport, energy or defence.

The improvement in weather forecasting throughout the second half of the 20$^{th}$ Century has been intrinsically linked to improvements in supercomputing. Weather simulation was one of the main applications that motivated the development of the first general use computer, the ENIAC, which emerged in the 1940s from work led by Von Neuman at the Princeton Institute for Advanced Study (Charney et al., 1950). From that moment, High Performance Computers (HPC) dedicated to weather forecasting have been some of the most powerful in the world and the computing capacity available to NMS has increased various orders of magnitude following Moore's Law[2], enabling increases in model complexity and resolution. However, there are good reasons to believe that we may have reached the end of this particular road.

---

[1] See Nate Silver's New York Times article "The Weather Man Is Not a Moron": https://www.nytimes.com/2012/09/09/magazine/the-weatherman-is-not-a-moron.html

A physical limit on the number of transistors that can be packed into silicon chips appears to have been reached, bringing an end to Moore's Law (Eli et al. 2017). This means that individual processors are not getting faster as rapidly as we have experienced in the past. In order to compensate for this, HPC architectures instead require an ever larger number of processors to increase their computational capacity. This adds engineering complexity and energy consumption, severely limiting the affordability of modern HPCs. Adapting Numerical Weather Prediction (NWP) models to HPCs with ever-increasing number of processors is extremely complex because of the overhead of sharing data between nodes and related scalability issues of the the underpinning scientific numerical algorithms (Adams et al., 2019).

Additionally, the last few decades have also seen a major increase in data volumes[3], so much so, that the data flowing from HPCs has now outstripped the downstream processing infrastructure. As a result, it is now difficult to complete scientific research and create downstream value-added services.

This is challenging for NMS, putting under pressure their traditional value-creation approach - which can succinctly be described as:

---

[2] Intel co-founder Gordon Moore's estimation in the 1960s that computing power would double every 18 months.

[3] Data continues increasing despite the end of Moore's Law: computers are not getting faster but there are more of them, also there is an ever growing number of sensors.

*improved HPC capabilities -> improved research and increased model resolution/complexity -> improved forecast skill -> increased socio-economic benefits.*

The context in which NMSs operate has also evolved substantially. From a political angle, the majority of NMS remain government agencies because forecast services are considered as "public good" due to the impact of weather forecast on population and infrastructure, and their implications for defence and the wider economy (over £30bn for UK alone, Heys et al., 2015). Internationally, NMS are part of the United Nations World Meteorological Organisation and share a long history of collaboration, including routine exchange of observations and forecasts. However, the rise of nationalistic agendas, and the increase of barriers-to-export on technologies underpinning weather forecasting (e.g. satellites, processors) may challenge international collaboration.

In various countries (USA, Netherlands) commercial exploitation of weather forecasts has been explicitly transferred to the private sector. More generally, economic pressures on public spending precipitated by the 2008 economic crisis have not only limited the funding available to NMS but increased the requirements on them to support growth of the broader economy on top of their traditional mission to protect life and property.

The way the public consumes weather is also changing. Weather information attracts big audiences from the general public (e.g. BBC Weather has over 70 million visits per month and the most frequent question to Amazon's Alexa is "what is the weather forecast"[4]). However, there has been a shift away from general broadcasting to personalised informa-tion via Apps or Digital Assistants, with an increasing demand for impact information (e.g. from "will it rain?" to "will my home be flooded?").

---

[4] Personal communication from Met Office and Amazon. Figures valid for December 2018

Raised environmental awareness in private business and government, together with the increasing cost and impact of extreme weather events, has raised demand for NMS services for decision-making. There is also an emerging demand for weather analysis for legal cases trying to attribute costs and responsibilities (e.g. case of the Urgenda Foundation vs The State of the Netherlands in 2015); and new laws and directives, such as EU INSPIRE, are emerging around publicly funded data access, locality and privacy. These needs require NMS to create a multiplicity of tools/products/platforms which let non-specialist consumers access complex information hidden in the forecast data.

Finally, technological discontinuities have brought new competitors from the technology trajectory: IBM bought Weather Channel Co. in 2015, reportedly for over $2bn[5]; a division of Panasonic developed forecasts for aviation; and Google's DeepMind has worked with National Grid on predicting energy demand.

In this paper we argue that the combination of contextual shifts and technological discontinuities present an unprecedented period of change that could substantially alter the foundations of the weather industry. Research shows that organisations are more likely to survive a recession than to adapt to a technological discontinuity (Anderson and Tushman, 1991). Therefore, current conditions are an existential threat to the weather industry. NMS need to evolve rapidly.

---

[5] https://www.wsj.com/articles/ibm-to-buy-weather-co-s-digital-data-assets-1446039939

Section 2 presents an analysis of current and emerging technologies, using the Technology Lifecycle Model as a diagnostic tool. Section 3 presents an analysis of structural vulnerabilities in traditional NMS value-creation models. Finally, Section 4 presents a series of recommendations to enable NMS to adapt to the new situation.

**The National Meteorological Services' dilemma**

The origins of weather forecasting are deeply rooted in government ownership and scientific research. Unsurprisingly, a clear signature of these origins remains in the current organisations and industry dynamics. NMS have evolved into complex, 24/7 operational organisations underpinned by sophisticated technology but they are still pursuing the same science-led value creation strategy adopted in the 1950s. In its simplest form this value creation strategy can be summarised as:

*improved HPC capabilities underpin improved research and enable increased model resolution and complexity. This leads to improved forecast skill which is then converted into increased socio-economic benefits.*

It is worth noting that this way of thinking about value creation is reinforced in NMS by prevailing organisational social structures which are dominated by technology and science, and operational structures which prioritise resilience and robustness of service within highly vertically integrated products, a natural consequence of supporting critical activities such as aviation and emergency response. However, the technological discontinuities and contextual changes described in the introduction bring these assumptions into question.

A useful tool to analyse the current situation is the Technology Lifecycle Model (Tushman and Anderson, 2004), a model which is not dissimilar to the scientific discovery lifecycle postulated earlier by Kuhn (1962). The Technology Lifecycle Model consists of four phases:

1) Once a consensus has emerged around a particular technology, complementary components and concepts emerge, giving rise to a *Dominant Design*.

2) This is followed by an *Era of Incremental Innovation*, during which routine, customer-led improvements take place. The exploitation of existing technologies during the period of incremental innovation becomes harder with time until …

3) a *Technological Discontinuity* is reached and new technologies emerge - often, initially offering lower performance than existing technologies. The new technologies open…

4) an *Era of Ferment*, characterised by high uncertainty and driven by the experimentation of those developing the new technologies, not by customer demand. Over time, the various unknowns and competing aspects of new technologies are resolved, interdependencies and integration emerge, and consensus around a new *Dominant Design* arises, starting a new iteration of the cycle.

Whereas other industries have gone through various iterations of this cycle, it is the first time for the weather industry: once the mathematical foundations for weather forecasting were established, an *Era of Ferment* took place between the 1920s-1940s, in which the likes of Lewis Fry Richardson developed the principles of distributed numerical weather simulation. This was rapidly followed by the emergence in the 1950s of a *Dominant Design,* which included the underpinning use of computers and established the major components of weather forecasting. From the 1960s until the present we have been immersed in an *Era of Incremental Innovation*[6], during which, underpinned by continuously increasing computational capacity, all elements of the *Dominant Design* for weather forecasting were improved upon.

---

[6] The last sixty years of weather forecasting are, no doubt, the story of a remarkable scientific success. The term is not intended to diminish the value of the many outstanding scientific and technological contributions but, simply, to reflect that those changes have taken place as part of an established dominant design.

The death of Moore's Law and the emergence of new technologies such as scalable cloud compute and machine learning have created a major *Technological Discontinuity*, opening a new *Era of Ferment* for weather forecasting. Clear evidence of this can be seen in the fact that, within the current trajectory for NWP, much higher investment and resources are now required than in the past in order to achieve similar (or smaller) increases in performance. Before analysing the new technologies and their role in the *Era of Ferment* that we are entering, it is worth to further expand on the issue around high data volumes.

Major NWP centres generate hundreds of terabytes per day. This is an enormous amount of data. As an illustration, the Met Office's archive is expected to reach one exabyte - $10^9$ gigabytes or approximately the equivalent of 4 billion high-definition films - before 2020. To put it simply, our ability to generate data (a direct consequence of the successful parallelisation and optimisation of dynamical models on HPC until now) has vastly outstripped our capacity to analyse data downstream. It should be noted that the process of analysing data, because of its ad hoc nature, is much more difficult to standardise and optimise than the process of generating data.

High data volumes slow down and complicate the process of scientific research and analysis. This has several pathological effects (Robinson et al 2019): we optimise the utilisation of data processing hardware at the cost of making inefficient use of our expensive experts (scientists and forecasters) who spend a substantial amount of their time dealing with data management and waiting for analyses to run on overcrowded systems. This creates pressures for analysts to make pragmatic choices, for instance subsetting or averaging data based on a priori assumptions, increasing the risk of confirmation bias. Finally, scientific creativity itself suffers when its workflows are regularly interrupted by inadequate systems (Speier et al., 1999; Hwang and Lin, 1999)

Modern data volumes also affect downstream exploitation by end-users. Weather forecasts are perishable (a forecast for tomorrow is more valuable today than it will be tomorrow) and moving and analysing such data volumes in a timely fashion is no longer feasible with traditional IT infrastructures. End-users have traditionally dealt with this problem in a similar way to scientists: by reducing the data; for example, by calculating the most likely weather symbol over a specific location at a given time. However, that means a substantial loss of information content because the nature of weather information itself has become more complex.

Increasingly, the skill we add with every new HPC generation does not simply manifest as more accurate predictions of the same old products. This is largely because we are now at the stage of resolving processes with a much finer chaotic scale (e.g. convection), and our skill is inherently probabilistic, rendering it more difficult to interpret and use. As we create more complex information, we increasingly need better ways to use it within sophisticated decision-making systems requiring full knowledge of the probabilities of upcoming weather events and access to non-environmental data. It is also the case that for events such as convection or tornadoes the time window to drive decision-making systems may be extremely short.

New technologies - principally scalable cloud compute (CC) and machine learning (ML) but also others such as Virtual/Augmented Reality and Natural Language Processing - are crucial to the new approach required within the *Era of Ferment* we are entering. In this paper we will focus on CC and ML which can be described as follows.

Cloud computing (CC) is the elastic provision - as much or as little as required - of on demand storage, network, and computational resources. Cloud technologies emerged as a by-product of the large IT infrastructure required by companies such as Amazon, Microsoft, and Google to deliver their own services and products. Built to support their peak-demand, these companies found that they had surplus infrastructure that they could monetise. It has now evolved into a mature business, with revenues estimated by Gartner in excess of $200bn in 2017. In effect, many different consumers share the same pool of resources, on the basis that their peak demand are vanishingly unlikely to coincide, conferring elasticity from the consumers' point of view, and efficient utilisation from the cloud suppliers point of view.

Machine learning (ML) can be described as a type of computer algorithm that can learn from data. As such, it is not a fixed computer code that has been programmed to solve a particular task - as, for example, the codes that calculate solutions to a set of equations for weather prediction do - but one that represents a statistical model and contains modifiable parameters that the computer code can adjust itself in order to achieve the best performance for a particular criterion and dataset (Alpaydin, 2016).

The elastic nature of CC makes possible the creation of scalable data science platforms that address the above-mentioned bottlenecks for scientific research and downstream value creation. Platforms such as Pangeo[7] (Robinson et al., 2019) have emerged in the last couple of years (with funding from US National Science Foundation and NASA, amongst others) and have already become active open communities involving many world-leading research institutes in geosciences (National Center for Atmospheric Research, Columbia University and Met Office to name a few). Platforms such as this facilitate the combination of NWP data with other datasets (transport, energy, or retail data) and the completion of sophisticated mathematical operations (including ML) without having to reduce information a priori. The application of ML to environmental sciences is still in an early stage but it has already proven its usefulness (Karpatne et al., 2019). For example, successful energy demand forecasts or weather predictions 24 hours ahead have been demonstrated (Grover et al. 2015).

---

[7] www.pangeo.io

As is typically observed in *Eras of Ferment,* the performance of CC- and ML-based technologies for weather forecasting is, currently, often below the performance offered by existing technologies. However, as demonstrated in many other industries, it would be wrong to assume that this could not change rapidly (Anderson and Tushman, 1991). In addition, the combination of CC and ML creates new opportunities, reducing the barriers of entry to those outside the traditional weather industry (see, for example, https://salientpredictions.launchrock.com/ and https://www.climacell.co/). Therefore, competition will increase, likely further accelerating improvements from new technologies.

Thus, as expected in this phase of fermentation in the Technology Lifecycle Model, NMS face a high level of uncertainty, with potentially conflicting short- and long-term risks and investment needs. This situation forces upon NMS the well known Innovator's Dilemma (Christensen, 1997): to either *Exploit* existing technologies (continue using resources to bring diminishing improvements and revenues now); or *Explore* new technologies and approaches (using resources instead to innovate and generate value in the future).

The combination of diminishing returns from existing technologies and the rapidly changing context indicate the need for NMS to *Explore*. However, as found by Christensen (1997), established organisations tend to focus efforts on the *Exploitation* of existing tech-nologies. This tendency intensifies further when organisations are under pressure because resource allocation is often shaped by decision-makers who are tied to previous choices and trying to satisfy existing customers while immersed within a culture unsupportive of different opinions and often unaware of the wider situation (Christensen and Bower, 1996). This is likely to be the situation at most NMS.

Importantly, new technologies also bring new requirements, challenging existing power and status structures. In other words, an *Era of Ferment* consist of social competition as much as of technological competition and it should be expected that new approaches will not be welcome by incumbents.

NMS may find it difficult to accept the required changes, not only because the current *Dominant Design* has been in place for many decades but because of the roots and weight of scientific research in the industry. As described by Kuhn - who identified the concept of paradigm shift in his work "The structure of scientific revolutions" (1962) - bringing change to mature scientific disciplines is extremely difficult because "scientific research is directed to the articulation of those … theories that the paradigm already supplies". As a consequence, "scientists do not normally aim to invent new theories and they are often intolerant of those invented by others".

The relevance of the social aspects of change should not be underestimated. It has been found across industries that, because of the complexities of the social element, it is often the case that executive leadership postpones transformation until a financial crisis is hit. As a result, in approximately 80% of cases the required change is coupled with CEO succession (Tushman et al., 1986). In the particular case of research-oriented organisations, as recently demonstrated by Wu et al. (2019), large entities tend to be more incremental and less innovative than smaller research teams. Additionally, NMS are also part of a global community (WMO) which intensifies pressures for incremental innovation.

This analysis shows that, when faced with the Innovator's Dilemma, NMS are likely to experience pressure to fall into the trap of focusing resources on *Exploit* only. Given the current shifts in context and technologies, that approach would be a major risk. NMS need to simultaneously pursue both *Exploitation* and *Exploration*.

**Strategic Analysis**

As concluded in previous sections, the current *Era of Ferment* requires the simultaneous pursuit of *Exploitation* (of existing technologies) and *Exploration* (of new technologies). It is worth making explicit that an *Era of Ferment* can be a long period. Agarwal and Bayus (2002) estimated that the average time from invention to commercialisation is approximately 30 years, timescales that are consistent with earlier experiences in the weather forecasting industry: it took 30 years from Richardson's proposal for NWP to the first real-time weather forecast. Therefore, NMS may be facing a long period of change.

However, as discussed in Section 2, there are forces driving NMS to focus their resources on *Exploit*, a strategy that carries substantial risks - trying to adapt to discontinuity through incremental adjustment has been shown to commonly fail (Tushman and O'Reilly, 1996) - and NMS need to recognise that successfully completing scientific research in weather/climate is not a valid substitute for the wider strategic innovation required.

To better understand how NMS could simultaneously *Exploit* and *Explore*, it is useful to separate the potential role of new technologies such as CC and ML as *competence-enhancing* (i.e. improving existing approaches/technologies) and *competence-destroying* (i.e. replacing existing approaches/technologies). It is also useful to separate the role of *Product* innovation (the generation of weather forecasts themselves) and downstream *Process* innovation (the combination of new and existing technologies to make the *Product* more valuable to users).

New technologies can be seen as *competence-enhancing,* complementing existing technologies and improving, particularly, the downstream *Process* of value-creation. One challenge is that some of the complementary assets which can enable *Process* innovation are increasingly owned by technology companies (Google, Amazon) or media conglomerates (Weather Channel, BBC). This is threatening to NMS because *Process* innovation, especially if driven by external players, can destroy the usefulness of the organisational knowledge embedded in the current structure and processes of NMS (Henderson and Clark, 1990). NMS should aim to ameliorate this risk and *Exploit* should be seen as an opportunity to rebalance NMS' focus and resources from *Product* to *Process* innovation (i.e. from the creation of forecasts to ensuring that forecasts are useful to end-users).

In addition, weather forecasts - the *Product* - have become a commodity[8] and in commodity markets competition focuses on price (Tushman and O'Reilly, 1996). This not only increases the existing financial pressures for NMS but accentuates their vulnerability to new entrants from the technology trajectory, who can be more agile and better at *Process* innovation. This is especially the case in a situation of slow product improvement as expected at the end of an *Era of Incremental Innovation.* The conclusion is that, even if new technologies such as CC/ML are considered as *competence-enhancing* to existing technologies, it is unlikely that NMS will be able to use them as effectively as their competitors to enter new markets. However, NMS are likely to remain important in highly regulated industries (aviation, defence) where they can exploit their existing value chain advantages.

New technologies can also be *competence-destroying,* enabling disruptive or radical changes, including at *Product* level. The most obvious example in the case of NMS would be the use of ML and CC to generate weather forecasts instead of solving physical equations using an HPC. This is, in fact, being tried and promising results using ML and CC have been demonstrated for short-range weather prediction 24 hours ahead (Grover et al., 2015; Weyn et al., 2019). This is dangerous for NMS as it overturns the existing value chains and attacks the foundations of the weather forecasting industry (Anderson and Tushman, 1991).

---

[8] undifferentiated and highly available. For example, all real-time forecasts from the National Centre for Environmental Prediction (USA) are free.

In summary, the technological discontinuities we are seeing can destroy existing competences and create an inflection point, substantially changing the underlying structure of the weather industry. In these conditions, the traditional value creation of NMS is unlikely to remain valid for long and NMS need to become actively involved in the shaping the next Dominant Design and their new value proposition. Given the size and resources of the new competitors from the technology trajectory NMS should not underestimate the threat.

An *Era of Ferment* requires managing high uncertainty and risk but offers an excellent opportunity to redefine partnerships and value creation strategies. Crucially, the simultaneous pursuit of *Exploit/Explore* needs to be considered beyond the limits of the individual NMS and requires including an extended range of actors. This is partly because of the limited knowledge of new technologies within NMS but also because NMS are no longer facing disruption from other NMS. The competition has now extended to organisations in the technological trajectory who are also present in the verticals of the value chain (e.g. Google, Microsoft); incumbents in proximate markets such as Panasonic or IBM; and, thanks to the reduced barriers to entry from data availability, compute resources and knowledge, to unexpected new competitors (including graduates students with expertise in ML or existing service providers to specific sectors).

The creation of a new ecosystem will require that NMS increase the numbers of staff with expertise in these new technologies because only they will have the legitimacy to manage the rapidly changing interdependencies and ensure access to partners and knowledge.

This latter point needs to be highlighted because, in an *Era of Ferment*, traditional tools such as market positioning or segmentation will not be useful tools to decide what the required balance of investment between *Exploit* and *Explore* are - demand is inchoate (Geroski, 2003) and change is driven by those developing the technologies.

**Conclusions and Recommendations**

The weather forecasting industry has entered a major period of transformation, an *Era of Ferment,* in which NMS need to redefine how they will create value as part of a new *Dominant Design*.

This is challenging. Our *Product* - the weather forecast - has not only been commoditised but it is also at risk of radical innovation by new entrants that could destroy the current competences of NMS. It is also an opportunity, NMS have substantial scientific and technical expertise that could be leveraged to create new partnerships and develop novel value creation strategies. However, NMS must take an active role in the discovery of the new *Dominant Design* and, as our analysis shows, simultaneously pursue Exploitation within the current technology trajectory and Exploration of new technologies in order to define a new value proposition.

In the case of *Exploit*, NMS should rebalance their focus and resources towards *Process* innovation, using their domain expertise as a leverage. This is particularly important because the data we are dealing with is inherently and increasingly complex (e.g. multi-variable, high-resolution probabilistic environmental predictions) and new entrants typically lack the expertise to exploit it optimally. It should be noted that the organisational inertia within NMS is likely to be substantial but, if focus and resources are not rebalanced, NMS will carry a high risk.

In the case of *Explore*, a new *Dominant Design* which could substantially affect the underlying structure of the industry is likely to emerge. NMS will not be able to afford the necessary resources to pursue exploration alone and, therefore, they should prioritise the creation of a unique ecosystem around them - including academia, technology companies, start-ups and others in the wider weather industry and beyond. NMS will need to balance competition and collaboration activities adequately.

The following conclusions and recommendations are intended as a common starting point for further thinking and reflection across different National Meteorological Services.

## 1. Simultaneous pursuit of Exploitation and Exploration

Exploit is efficiency-oriented with formalised roles and processes designed to capture existing opportunities; Explore is experimentation-oriented with looser and more informal structures grouping heterogeneous skills to create new opportunities (Tushman and Smith, 2004).

This requires different structures and organisational changes that should also support the dynamic reconfiguration of resources (Teece et al., 1997). Fundamentally, this requires a profound re-evaluation of the full activity system at individual NMS including their people, culture and structures (Tushman and O'Reilly, 1997). The new situation needs people with increased agency and flexibility, including managers who are less process-oriented and more able to manage relationships outside their teams. This demands a culture of openness and informed risk-taking.

## 2. Supported by an emergent new culture

There is a necessary transition for organisations that have successfully evolved through seventy years of incremental improvement; current habits and mundane actions embedded into NMS are a direct consequence of that period. As Chambliss (1989) identified, it is the alignment of many small mundane activities and skills that is at the source of excellence. Because of this, transforming NMS will be much more difficult than just introducing a new organogram or process (Tushman and O'Reilly, 1996). Organisational inertia, as well as conscious and unconscious resistance by incumbents are very real issues that need to be acknowledged. At the same time, in most organisations there is often a pent-up demand for change that tends not to be recognised at senior levels (Tushman et al., 1986). Therefore, the emergence of a new culture requires time and staff involvement: it needs to be a process of dialogue and co-creation. The opposite, a manufactured new culture developed by external consultants and executives behind closed doors, followed by a communication campaign is unlikely to succeed.

**3. Enabled by an underpinning, unifying strategy**

The simultaneous pursuit of *Exploit/Explore* (with its required parallel and different structures) and the emergence of a new culture supporting it require from each individual NMS the clear articulation of what their purpose is, defining an underpinning and unifying strategy.

A strategy should be understood as a "theory of value creation" (Zenger, 2013). In other words, each individual NMS should understand how it creates value by combining its unique resources and capabilities with other external assets. Crucially, it should be a continuously updated theory (i.e. not a static 5 year plan) which not only provides long-term direction but a clear rationale to manage daily decision-making, particularly around horizontal and vertical boundaries internally (within and across Business Units) and externally (with suppliers, customers, and competitors).

As our analysis shows, a critical aspect of the new situation facing NMS is that it requires them to develop ecosystems (i.e. networks of partners, suppliers, collaborators) to pursue the exploration of new technologies and, particularly, *Process* innovation. The extent to which each NMS balances resources and the way it interfaces with external organisations has to be determined individually. However, it needs to be recognised that traditional, efficiency-oriented, transactional arrangements that are successful for *Exploitation* are less suitable for *Exploration*.

Finally, it should be acknowledged that a strategy enabling *Exploitation and Exploration* should create the right conditions to facilitate innovation as an emerging property across the organisation, not to dictate to each individual department how innovation should be done. This puts a clear demand on senior leaders: aligning conflicting aims and reconfiguring assets and skills is their critical function to enable the organisation to sustain competitive advantage (O'Reilly and Tushman, 2008).

The recommendations above indicate that we should accept that articulating what value the NMS creates and how it creates it is a continuous process, requiring understanding of where we are in the Technology Lifecycle Model. Equally, leadership at NMS need to accept the cost of the heterogeneous organisational structures and mechanisms required to solve the inherent conflicts and contradictions that will appear. If this cost is not accepted NMS will be unable to facilitate the emergence of the new skills, habits and culture required in this period of transformation.

In the case of commercial organisations, failure to do so risks bankruptcy. In the case of government departments or agencies, it risks obsolescence.

# References

Abbe, C. (1901). The physical basis of long-range weather forecasts. *Mon. Weath. Rev.*, 29, 551-561.

Adams, S.V., Ford, R.W., Hambley, M., Hobson, J.M., Kavčič, I., Maynard, C.M., Melvin, T., Müller, E.H., Mullerworth, S., Porter, A.R., Rezny, M., Shipway, B.J., Wong, R. (2019). LFRic: Meeting the challenges of scalability and performance portability in Weather and Climate models. *Journal of Parallel and Distributed Computing*, 132, 383-396.

Agarwal, R., & Bayus, B. (2002). The market evolution and take-off of product innovations. *Review of Marketing Science*, 48(8), 1024-1041.

Alpaydin, E. (2016). *Machine Learning*. MIT Press.

Anderson P., & Tushman M. L. (1991). Managing though cycles of technological change. *Research Technology Management*, 34(3), 26–31.

Bauer, P., Thorpe, A., & Brunet, G. (2015). The quiet revolution of numerical weather prediction. *Nature,* 525. 47-55.

Bjerknes, V. (1904). Das Problem der Wettervorhersage betrachtet vomStandpunkt der Mechanik und Physik. *Meteorol. Z.*, 21, 1-7.

Bolin, B. (1955). Numerical forecasting with the barotropic model. *Tellus*, 7, 27-49.

Bungay, S. (2011). *The art of action: How leaders close the gaps between plans, actions, and results*. Nicholas Brealey Publishing

Chambliss, D. F. (1989). *Champions : the making of Olympic swimmers.* Morrow.

Charney, J. G., Fioertoft, R. & Neumann, J. v. (1950). Numerical integration of the barotropic vorticity equation. *Tellus*, 2, 237-254.

Christensen, C. M. (1997). *The innovator's dilemma: when new technologies cause great firms to fail.* Harvard Business School Press

Christensen, C. M., & Bower, J. L. (1996). Customer power, strategic investment, and the failure of leading firms, *Strategic Management Journal*, 17(3), 197-218.

Elie, T., Forbes, N., & Strawn, G. (2017). The End of Moore's Law. *Computing in Science & Engineering*, 19(2), 4-6.

Foster, R. N. (1986). *Innovation: the attacker's advantage*. Summit Books.

Geroski, P. (2003). *The evolution of new markets.* Oxford University Press.

Grover, A., Kapoor, A. & Horvitz, E. (2015). A Deep Hybrid Model for Weather Forecasting. In Proceedings of the 21th ACM SIGKDD. *International Conference on Knowledge Discovery and Data Mining (KDD '15)*. ACM, New York.

Hansen, M. T., & Birkinshaw, J. (2007). The Innovation Value Chain. *Harvard Business Review*, 85(6), 121–130.

Hargadon, A., & Sutton, R. I. (2000). Building an Innovation Factory. *Harvard Business Review*, 78(3), 157–166.

Henderson, R., & Clark, K. (1990). Architectural innovation: The reconfiguration of existing product technologies and the failure of established firms. *Administrative Science Quarterly*, 35(1), 9-30.

Heys, R., Duke, C., Muller, P., Ladher, R. & Koch, L. (2015). *Met Office General Review: Economics Analysis*. London Economics.

Hwang, M.I. and Lin, J.W. (1999). Information dimension, information overload and decision quality. *Journal of Information Science*. 25, 213-218

Karpatne, A., Ebert-Uphoff, I., Ravela, S., Ali-Babaie H., & Kumar, V. (2019). Machine Learning for the Geosciences: Challenges and Opportunities *IEEE Transactions on Knowledge and Data Engineering*, 31(8).

Kuhn, T. S. (1962). *The structure of scientific revolutions*. University of Chicago Press.

Lorenz, E. N. (1963). Deterministic non-periodical flow. *J. Atmos. Sci.* 20, 130-141.

Nadler, D. A., & Tushman, M. L. (1997). *Competing by design: The power of organisational architecture.* Oxford University Press.

O'Reilly, C. A. & Tushman, M. L. (2008). Ambidexterity as a dynamic capability: Resolving the innovator's dilemma. *Research in Organizational Behavior*, 28, 185-206.

Porter, M. E. (1980). *Competitive Strategy.* Free Press.

Richardson, L. F. (1922). *Weather prediction by numerical process*. Cambridge Univ. Press.

Robinson, N. McCaie, T., Tomlinson, J., Hilson, A., Powell, T., Prudden, R., & Arribas, A. (2019). *Giving Scientists back their flow: Analysing Big Geoscience Data-sets in the Cloud*. In press. American Geophysical Union Books.

Silver, N. (2012). *The signal and the noise: Why most predictions fail but some don't*. Penguin Press.

Speier, C. , Valacich, J. S. and Vessey, I. (1999). The Influence of Task Interruption on Individual Decision Making: An Information Overload Perspective. *Decision Sciences*, 30, 337-360.

Teece, D. J., Pisano, G., & Shuen, A. (1997). Dynamic capabilities and strategic management. *Strategic Management Journal*, 18(7), 509-533.

Tushman M. L. & Anderson P. (2004). *Managing strategic innovation and change: a collection of readings*. Oxford University Press.

Tushman, M. L., Newman, W. H., & Romanelli, E. (1986). Convergence and upheaval: Managing the unsteady pace of organizational evolution. *California Management Review*, 29(1), 29-44.

Tushman, M. L., & O'Reilly, C. A. (1996). Ambidextrous organizations: managing evolutionary and revolutionary change. *California Management Review*, 38(4), 8–29.

Tushman, M., & O'Reilly, C. A. (1997). *Winning through Innovation: A Practical Guide to Leading Organizational Change and Renewal.* Harvard Business School Press.

Tushman, M., & Smith, W. K. (2004). *Innovation streams, organization designs, and organizational evolution*. In: Tushman M. L. & Anderson P. (2004). *Managing strategic innovation and change: a collection of readings*. Oxford University Press, 2-17.

Ventresca, M., & Levin, P. (2004). Information and ambiguity in markets. *Accounts: A Newsletter of Economic Sociology*, 5/1. pp. 1-3

Wageman R., & Hackman, J. R. (2010). *What makes teams of leaders leadable?* In: Nohria, N., & Khurana, R. (Eds.). (2010). *Handbook of leadership theory and practice*, Harvard Business Review Press, 475-506.

Weyn, J. A., Durran, D. R., & Caruana, R. (2019). Can machines learn to predict weather? Using deep learning to predict gridded 500-hPa geopotential height from historical weather data. Journal of Advances in Modelling Earth Systems,11.

Zenger, T. (2013). What is the theory of your firm. *Harvard Business Review, 91(6), 72–78.*